\documentstyle[aps,epsfig]{revtex}
\begin{document}

\draft

\title{Superconducting phase transition in the Nambu - Jona-Lasinio model}

\author{M. Sadzikowski}

\address{Institute of Nuclear Physics, Radzikowskiego 152, 31-342 Krak\'ow, Poland}

\maketitle

\tighten

\begin{abstract}
The Nambu - Bogoliubov - de Gennes method is applied to the problem of superconducting QCD. 
The effective quark-quark interaction is described within the framework of the Nambu - Jona-Lasinio
model. The details of the phase diagram are given as a function of the strength of the quark-quark
coupling constant $G^{\,\prime }$. It is find that there is no
superconducting phase transition when one uses the relation between the coupling constants $G^{\,\prime }$ and 
$G$ of the Nambu - Jona-Lasinio model which follows from the Fierz transformation. However, for
other values of $G^{\,\prime }$ one can find a rich phase structure containing both the chiral and the superconducting
phase transitions.
  
\end{abstract}


\section{Introduction}

\bigskip

In this letter we apply the Nambu - Bogoliubov - de Gennes (NBG) \cite{NBG} method to the problem
of superconducting QCD \cite{cs,cs2}. The Nambu - Jona-Lasinio hamiltonian for two flavors and three colors \cite{njl} 
is chosen as a model for the quark interaction:
\begin{equation}
\label{njl}
H = \int d^3x\left\{ \bar{\psi}\left( -i\vec{\gamma }\cdot\vec{\nabla } + m\right)\psi - 
G\left[ \left(\bar{\psi}\psi\right)^2 + \left(\bar{\psi}i\gamma_5\vec{\tau }\psi\right)^2\right]\right\} ,
\end{equation}
where $m$ is a current quark mass, $\vec{\tau }$ is the vector of Pauli matrices, and all indices are suppressed.
The hamiltonian (\ref{njl}) describes the quark interaction. It works under the assumption
that the deconfinement phase transition appears at a lower baryon density then the chiral phase 
transition. In that case we can think of quarks as relevant degrees of freedom at appropriate densities. 

The NBG method is a very convenient tool for mean-field calculations of any hamiltonian 
containing four-fermion interactions. It allows us to trace the physical details of the considered problem
and is easy to apply. The problem of the superconducting QCD within the framework of Nambu - Jona-Lasinio
model has been considered before in \cite{klev_papp}. This letter differs in the approach and
in details of the regularization procedure. Instead of using the three-momentum cut-off for all
interactions we choose the form-factor regularization for diquark interactions and leave the cut-off
only to regularize the infinite contribution from the Dirac sea of antiquarks. The cut-off regularization
influences only the features of the chiral phase transition and has only minor meaning for the superconducting
phase transition.

In sections II and III we describe the Fierz transformation and the mean-field approximation.
Section IV gives the details of NBG approach. The last section contains numerical results.

\section{Fierz transformation}

Through the use of appropriate Fierz transformation one can rearrange the hamiltonian (\ref{njl}) 
to the form which contains color {\bf 6} and $
\bar{{\bf 3}} $ operators explicitly. To this end, one can use the relations between Pauli matrices:

\begin{eqnarray} 
&& \vec{\tau }_{ik}\vec{\tau }_{lj} + \delta_{ik}\delta_{lj} = -(\tau_2)_{il}(\tau_2)_{jk}+ 
(\vec{\tau}\tau_2)_{il}(\tau_2\vec{\tau })_{jk} \nonumber \\
&&\vec{\tau }_{ik}\vec{\tau }_{lj} - \delta_{ik}\delta_{lj} = -2(\tau_2)_{il}(\tau_2)_{jk} 
\end{eqnarray} 
and the following decomposition of unity in the $SU(3)$ color group:
\begin{equation} 
\delta_{\alpha\beta }\delta_{\gamma\delta } = \frac{1}{3}\delta_{\alpha\gamma}\delta_{\delta\beta }+
2 \left(\frac{\lambda^a}{2}\right)_{\alpha\gamma }\left(\frac{\lambda^a}{2}\right)_{\delta\beta } .
\end{equation}
From the Fierz transformed form of (\ref{njl}) one can verify that the interactions in the scalar or axial-vector 
$ \bar{{\bf 3}} $ and tensor or vector {\bf 6} channels are attractive, whereas in pseudoscalar or
vector $ \bar{{\bf 3}} $ and axial-vector {\bf 6} channels are repulsive. Assuming that only scalar $ \bar{{\bf 3}} $ creates condensate one
obtains the relation $G^{\,\prime } = G/4 $ between the coupling constant $G^{\,\prime }$ of the diquark scalar channel and the
coupling constant $G$ of the Nambu - Jona-Lasinio hamiltonian (\ref{njl}). Nevertheless in a more general treatment one
can try to use $G^{\,\prime }$ as an additional parameter, as we do in this letter.
Thus our starting point is the following hamiltonian operator:
\begin{eqnarray}
\label{h}
H = \int d^3x\left[ \bar{\psi}^{j}_{\alpha }\left( -i\vec{\gamma }\cdot\vec{\nabla }+m \right)
\psi^{j}_{\alpha } - G\left(\bar{\psi}^{j}_{\alpha }\psi^{j}_{\alpha }\right)^2-\right. \nonumber \\ 
\left. G^{\,\prime}\left(\psi^{j}_{\alpha }(\tau_2)_{jk}(t^A)_{\alpha\beta }C\gamma_5\psi^{k}_{\beta }\right)^\dagger
\left(\psi^{m}_{\delta }(\tau_2)_{mn}(t^A)_{\delta\rho }C\gamma_5\psi^{n}_{\rho }\right)\right] ,
\end{eqnarray}
where $j,k,m,n=u,d$ are flavor, $\alpha ,\beta ,\delta ,\rho =R,G,B$ are color indices and
spinor indices are suppressed. The quantity $\tau_2 $ is
antisymmetric Pauli matrix, $t^A$ is vector of three antisymmetric Gell-Mann matrices and $C=i\gamma_2\gamma_0$
is charge conjugation matrix. For $G, G^{\,\prime} >0$ hamiltonian (\ref{h}) describes the attraction of quarks in
scalar color singlet and scalar color antitriplet channels.

\section{Mean-field analysis}

To study the finite density effects at zero temperature we have to consider the
effective hamiltonian operator, $F=H-\mu N$, where $\mu $ is teh quark chemical potential 
and $N$ is particle number operator:
\begin{equation}
N=\int d^3x (\psi^{j}_{\alpha })^\dagger\psi^{j}_{\alpha } .
\end{equation}
The non-zero density of quarks influences the structure of the vacuum by developing
the energy eigenstates ladder of the Fermi sea quarks. However, as is well known, the Fermi sea is unstable in the presence
of the attractive channels leading to the Cooper instability. Thus we expect the creation of uniform condensates
in color singlet and antitriplet channels. The  non-zero color singlet condensate, called the chiral condensate, 
breaks the chiral symmetry and leads to the phenomenon of mass generation for quarks \cite{njl}. 
The non-zero color antitriplet condensate breaks the color gauge symmetry and
leads to the phenomenon of Colour Superconductivity \cite{cs,cs2}. Both phenomena can be described in the mean-field approximation. 
Assuming that the fluctuations of the fields around the uniform condensates are small we linearize the four-fermion
operators:
\begin{eqnarray}
&& \left(\psi\tau_2t^AC\gamma_5\psi\right)^\dagger\left(\psi\tau_2t^AC\gamma_5\psi\right)\approx
\left(\psi\tau_2t^AC\gamma_5\psi\right)^\dagger\langle\psi\tau_2t^AC\gamma_5\psi\rangle+ \nonumber\\
&& \langle\left(\psi\tau_2t^AC\gamma_5\psi\right)^\dagger\rangle\psi\tau_2t^AC\gamma_5\psi-
\langle\left(\psi\tau_2t^AC\gamma_5\psi\right)^\dagger\rangle\langle\psi\tau_2t^AC\gamma_5\psi\rangle ,
\end{eqnarray}
and
\begin{eqnarray}
(\bar{\psi}\psi)^2\approx 2\langle\bar{\psi}\psi\rangle\bar{\psi}\psi - \langle\bar{\psi}\psi\rangle^2 .
\end{eqnarray}
The brackets in the above formulae denotes the vacuum expectation values of the field operators. Assuming that
the lowest energy state is uniform\footnote{This assumption is not obvious, see e.g. \cite{sb}} we introduce the chiral condensate $\bar{M}$:
\begin{equation}
\langle\bar{\psi}\psi\rangle = -\frac{\bar{M}}{2 G}
\end{equation}
and the superconducting condensate $\Delta $:
\begin{equation}
\langle\psi\tau_2t^AC\gamma_5\psi\rangle = \frac{\Delta}{2 G^{\,\prime }} \delta^{A1} ,
\end{equation}
where the direction in the color space is chosen arbitrarily. With the new variables, the
effective hamiltonian in the mean field approximation takes the form:
\begin{eqnarray}
\label{f}
F = \int d^3x\left[ \psi^\dagger\left( -i\vec{\alpha }\cdot\vec{\nabla }+M\gamma_0 -\mu\right)\psi -
\frac{\Delta }{2}\left(\psi\tau_2t^1C\gamma_5\psi\right)^\dagger - 
\frac{\Delta^\ast }{2}\psi\tau_2t^1C\gamma_5\psi +
\frac{|\Delta |^2}{4 G^{\,\prime }} + \frac{(M-m)^2}{4 G}\right] ,
\end{eqnarray}
where $M = \bar{M}+m$ is the constituent quark mass.

\section{Nambu-Bogoliubov-de Gennes approach}

We use the NBG approach
which is particularly useful in practice and enlights the physical aspects of the considered problem.
To start with, it is very convenient to introduce the creation and annihilation operators:
\begin{equation}\label{anni}
\psi^{j}_{\alpha }(t,\vec{x}) = \sum_{s=1,2}\int\frac{d^3k}{(2\pi )^3\sqrt{2 E_k}}
\left\{ u_s(\vec{k}) a^j_{\alpha ,\, s}(\vec{k}) \exp (-i k x)+ 
v_s(\vec{k}) b^{j\,\dagger }_{\alpha ,\, s} (\vec{k}) \exp (i k x)\right\} ,
\end{equation}
where $E_k=\sqrt{\vec{k}^2+M^2}$, $u_s, v_s$ are Dirac bispinors, $a^i_{\alpha ,\, s}(\vec{k})$ 
$(b^{i}_{\alpha ,\, s} (\vec{k}))$ is an
annihilation operator of quark (antiquark) of color $\alpha $, flavor $j$, spin $s$,
and momentum $\vec{k}$, satisfying anticommutation relations:
\begin{equation}
\left\{a^i_{\alpha ,\, s}(\vec{k}),a^{j\,\dagger }_{\beta ,\, r} (\vec{p})\right\}=
\left\{b^i_{\alpha ,\, s}(\vec{k}),b^{j\,\dagger }_{\beta ,\, r} (\vec{p})\right\}=
(2\pi )^3\delta_{sr}\delta_{\alpha\beta }\delta (\vec{k}-\vec{p}).
\end{equation}

The theories with point four-fermion interactions are not renormalizable and require 
regularization which introduces additional parameters. We use the
form-factor regularization, as well as the standard
three-momentum cut-off. The form-factor regularization introduces the function $f(\vec{p})$
for each fermi field in (\ref{f}) that cuts the large momenta in the interaction.
Following \cite{cs2} we use:
\begin{equation}
f(\vec{p})=\frac{\omega^2}{\omega^2+\vec{p}^2},
\end{equation}
where $\omega $ is of order of 0.4-1.0 GeV. The use of the form-factor  is a convenient regularization
of color superconducting part of (\ref{f}). However, in the description of chiral symmetry braking it unnecessarily complicates
the shape of the Fermi surface. Thus for that part we  use the three-momentum cut-off
regularization. This introduces another parameter, $\Lambda $. The parameters $G=5.01$ GeV$^{-2}$, 
$\Lambda = 0.65$ GeV are fixed within the Nambu - Jona-Lasinio model. They directly influence the features of the chiral 
symmetry breaking \cite{klev}. In the numerical calculations we put arbitrarily $\omega=0.8$ GeV.
The features of the chiral symmetry breaking are robust against the details of the 
form-factor regularization. However, such quantities as
the value of the gap parameter $\Delta $, or the exact position of the superconducting phase transition in the phase diagram 
do depend on these details. We discuss this point in the last chapter.
By using the decomposition (\ref{anni}) we can write the effective hamiltonian (\ref{f}) 
in the form\footnote{Time play no role in our considerations thus was chosen arbitrary to $t=0$.}:
\begin{eqnarray}
\label{f2}
&& F = \int\frac{d^3p}{(2\pi )^3}\left\{\sum_{s=\uparrow , \downarrow}
\left[ (E_p-\mu) a^{j\,\dagger }_{\alpha ,\, s}(\vec{p})a^{j}_{\alpha ,\, s}(\vec{p})-
(E_p+\mu) b^{j}_{\alpha ,\, s}(\vec{p})b^{j\,\dagger }_{\alpha ,\, s}(\vec{p})\right]\right. + \nonumber \\
&& \Delta^{\;\alpha\beta}_{\,jk}
a^{j\,\dagger }_{\alpha ,\, \downarrow}(\vec{p})a^{k\,\dagger }_{\beta ,\, \uparrow}(-\vec{p})-
\Delta^{\ast\,\alpha\beta}_{\,jk}
a^{j}_{\alpha ,\, \downarrow}(\vec{p})a^{k}_{\beta ,\, \uparrow}(-\vec{p}) + \nonumber \\
&& \Delta^{\;\alpha\beta}_{\,jk}
b^{j}_{\alpha ,\, \downarrow}(\vec{p})b^{k}_{\beta ,\, \uparrow}(-\vec{p})-
\Delta^{\ast\,\alpha\beta }_{\,jk}
b^{j\,\dagger }_{\alpha ,\, \downarrow}(\vec{p})b^{k\,\dagger }_{\beta ,\, \uparrow}(-\vec{p}) + \nonumber \\
&& \left. \frac{(M-m)^2}{4 G}+\frac{|\Delta |^2}{4 G^{\,\prime }} \right\} ,
\end{eqnarray}
where 
\begin{equation}
\Delta^{\;\alpha\beta }_{jk} = \Delta f(\vec{p})^2 (\tau_2)_{jk} (t^1)_{\alpha\beta } .
\end{equation}
The symbols $\uparrow ,\downarrow $ denote the spin projection of (anti)quarks on the direction $\vec{p}$. From the
formula (\ref{f2}) it follows that at the mean field level the quarks and antiquarks decouple
from each other and can be treated independently. Thus in the following we consider only the
quark part of the effective hamiltonian.
The first term and the second line of (\ref{f2}) describing the interaction of quarks can be write in the form:
\begin{equation}
\label{hpart}
F_{part}=\int\frac{d^3p}{(2\pi )^3}
\left(a^{j\,\dagger }_{\alpha ,\, \downarrow}(\vec{p}), a^{k}_{\beta ,\, \uparrow}(-\vec{p})\right)
\left(\begin{array}{cc}
E_p-\mu & \Delta^{\;\alpha\beta}_{\,jk}
 \\
\Delta^{\ast\,\alpha\beta}_{\,jk}
  & -(E_p-\mu) \\
\end{array}\right)
\left(\begin{array}{c}
a^{j}_{\alpha ,\, \downarrow}(\vec{p}) \\
 a^{k\,\dagger }_{\beta ,\, \uparrow}(-\vec{p}) \\
\end{array}\right) + 
2\cdot 3\int\frac{d^3pV}{(2\pi )^3}(E_p-\mu),
\end{equation}
where the last integral cancels the infinite contribution from the inverse ordering of creation and annihilation operators
in (\ref{hpart}) with respect to (\ref{f2}). The degeneracy factor $2\cdot 3$ counts flavors and colors.
The task is to diagonalize the hamiltonian $F_{part}$. It can be checked that (\ref{hpart}) 
can be written down as the sum over six independent families:
\begin{equation}
F_{part}=\int\frac{d^3p}{(2\pi )^3}\left[\sum_{a=1}^2 f^{a\,\dagger} A f^a+\sum_{a=3}^4 f^{a\,\dagger} B f^a +
\sum_{a=5}^6 f^{a\,\dagger} C f^a\right] + 6\int\frac{d^3pV}{(2\pi )^3}\epsilon
\end{equation}
where
\begin{equation}
A=\left(\begin{array}{cc}
\epsilon & -\Delta f(\vec{p})^2 \\
-\Delta^{\ast } f(\vec{p})^2 & -\epsilon \\
\end{array}\right) ,\;\;
B=\left(\begin{array}{cc}
\epsilon & \Delta f(\vec{p})^2 \\
\Delta^{\ast } f(\vec{p})^2 & -\epsilon \\
\end{array}\right) ,\;\;
C=\left(\begin{array}{cc}
\epsilon & 0 \\
0 & -\epsilon \\
\end{array}\right) 
\end{equation}
$\epsilon = E_p - \mu $ and the families are:
\begin{eqnarray}\label{fam}
f^1=\left(\begin{array}{c}
a_{R,\,\downarrow }^u (\vec{p})\\
a_{G,\,\uparrow }^{d\,\dagger } (-\vec{p})\\
\end{array}\right) ,\;\;
f^2=\left(\begin{array}{c}
a_{G,\,\downarrow }^d (\vec{p})\\
a_{R,\,\uparrow }^{u\,\dagger } (-\vec{p})\\
\end{array}\right) ,\;\;
f^3=\left(\begin{array}{c}
a_{G,\,\downarrow }^u (\vec{p})\\
a_{R,\,\uparrow }^{d\,\dagger } (-\vec{p})\\
\end{array}\right) ,\;\;\\\nonumber
f^4=\left(\begin{array}{c}
a_{R,\,\downarrow }^d (\vec{p})\\
a_{G,\,\uparrow }^{u\,\dagger } (-\vec{p})\\
\end{array}\right) ,\;\;
f^5=\left(\begin{array}{c}
a_{B,\,\downarrow }^u (\vec{p})\\
a_{B,\,\uparrow }^{u\,\dagger } (-\vec{p})\\
\end{array}\right) ,\;\;
f^6=\left(\begin{array}{c}
a_{B,\,\downarrow }^d (\vec{p})\\
a_{B,\,\uparrow }^{d\,\dagger } (-\vec{p})\\
\end{array}\right) .
\end{eqnarray}
The symbols $R,G,B$ denote colors and $u,d$ flavors quantum numbers of quarks. The last two families
describe the "Blue" quarks which are free and do not contribute to the color superconducting phase.
The diagonalization of $F_{part}$ is now straightforward because the problem has been reduced to the 
independent diagonalization of 2 by 2 hermitian matrices
$A$ and $B$. To this end one introduces the Bogoliubov unitary transformation of the vectors (\ref{fam}).
For families $a=1,2$ this transformation takes the form:
\begin{equation}\label{u}
U = \left(\begin{array}{cc}
\eta\sqrt{\frac{\epsilon + \lambda }{2 \lambda }} & \eta^\prime \sqrt{\frac{\lambda - \epsilon }{2 \lambda }}\exp (i\delta ) \\
-\eta\sqrt{\frac{\lambda - \epsilon}{2 \lambda }}\exp ( - i\delta ) & \eta^\prime\sqrt{\frac{\epsilon +\lambda }{2 \lambda }} \\
\end{array}\right) ,\;\;\lambda = \sqrt{\epsilon^2+|\Delta |^2 f(\vec{p})^4},\;\; \Delta = |\Delta |\exp (i\delta ),
\end{equation}
where $\eta , \eta^\prime $ are arbitrary phase factors.
For families $a=3,4$ Bogoliubov transformation is the same as (\ref{u}) but the change of the sign 
$\Delta\longrightarrow -\Delta $. Through the transformation (\ref{u}) one can introduce the quasiparticles basis:
\begin{equation}
\left(\begin{array}{c}
\alpha_1 (\vec{p})\\
\beta_{1}^{\dagger } (-\vec{p})\\
\end{array}\right) = U^\dagger
\left(\begin{array}{c}
a_{R,\,\downarrow }^u (\vec{p})\\
a_{G,\,\uparrow }^{d\,\dagger } (-\vec{p})\\
\end{array}\right)
\end{equation}
In the limit of vanishing interaction $|\Delta |\longrightarrow 0$, one can verify that:
\begin{equation}
\alpha_1(\vec{p}) =
\left\{\begin{array}{lc}
\eta a_{R,\,\downarrow }^u(\vec{p}) & \epsilon > 0 \\
\eta^\prime \exp (i\delta ) a_{G,\,\uparrow }^{d\,\dagger } (-\vec{p}) & \epsilon < 0 \\
\end{array}\right. ,\;\;
\beta_1(\vec{p}) = \left\{\begin{array}{lc}
\eta^\prime \exp a_{G,\,\uparrow }^{d} (-\vec{p}) & \epsilon > 0 \\
-\eta \exp (i\delta ) a_{R,\,\downarrow }^{u\,\dagger }(\vec{p}) & \epsilon < 0 \\
\end{array}\right. .
\end{equation}
The above equations confirm the interpretation of $\alpha^{\dagger}_1,\beta^{\dagger}_1$ as the creation
operators of particles above the Fermi surface and of holes below the Fermi surface.
The hole can be understood as a lack of particle from the occupied Fermi sea.
In the quasiparticles basis $F_{part}$ is diagonal and takes the form:
\begin{eqnarray}
&& F_{part}=\int\frac{d^3p}{(2\pi )^3}\sum_{a=1}^4\sqrt{\epsilon^2+|\Delta |^2 f(\vec{p})^4}
\left(\alpha_{a}^\dagger (\vec{p})\alpha_{a}(\vec{p})+\beta_{a}^\dagger (\vec{p})\beta_{a}(\vec{p})\right)+
\int\frac{d^3p}{(2\pi )^3}\sum_{s=\uparrow ,\downarrow }\sum_{j=u,d}\epsilon a_{B,s}^{j\,\dagger }(\vec{p})a_{B,s}^{j}(\vec{p})+ \nonumber \\
&& +\, 4\int\frac{d^3pV}{(2\pi )^3}\left(\epsilon -\sqrt{\epsilon^2+|\Delta |^2 f(\vec{p})^4}\right) .
\end{eqnarray}
The factor 4 in the last term comes from 2 flavors and 2 paired colors (except for the "Blue" one). 

The contribution from antiquarks can be find in exactly the same way. The only difference is
the Dirac sea contribution which is crucial for the chiral symmetry restoration (e.g. \cite{bubbala}).
Indeed  $b^{j}_\alpha $ operators in the first line of (\ref{f2}) have to be normally ordered. This ordering
leave the infinite Dirac sea contribution to the effective hamiltonian. The final expression for the full diagonal free
energy operator is:
\begin{eqnarray}
\label{fun}
&& F = \int\frac{d^3p}{(2\pi )^3}\sum_{a=1}^4\left\{\sqrt{(E_p-\mu)^2+|\Delta |^2 f(\vec{p})^4}
\left(\alpha_{a}^\dagger (\vec{p})\alpha_{a}(\vec{p})+\beta_{a}^\dagger (\vec{p})\beta_{a}(\vec{p})\right)\right. \nonumber \\
&& +\left. \sqrt{(E_p+\mu)^2+|\Delta |^2 f(\vec{p})^4}\left(\tilde{\alpha}_{a}^\dagger (\vec{p})\tilde{\alpha}_{a}(\vec{p})+
\tilde{\beta}_{a}^\dagger (\vec{p})\tilde{\beta}_{a}(\vec{p})\right)\right\} \nonumber \\
&& +\int\frac{d^3p}{(2\pi )^3}\sum_{s=\uparrow ,\downarrow }\sum_{j=u,d}\left\{
|E_p-\mu | c_{B,s}^{j\,\dagger }(\vec{p})c_{B,s}^{j}(\vec{p})+ 
(E_p+\mu )b_{B,s}^{j\,\dagger }(\vec{p})b_{B,s}^{j}(\vec{p})\right\} \nonumber \\
&& +\frac{M^2V}{4 G}+\frac{|\Delta |^2V}{4 G^{\,\prime }} \nonumber \\
&& +4\int\frac{d^3pV}{(2\pi )^3}\left\{ (E_p-\mu) - \sqrt{(E_p-\mu)^2+|\Delta |^2 f(\vec{p})^4}\right\} +
4\int\frac{d^3pV}{(2\pi )^3}\left\{ (E_p+\mu) - \sqrt{(E_p+\mu)^2+|\Delta |^2 f(\vec{p})^4}\right\} \nonumber \\
&& +4\int_{E_p<\mu }\frac{d^3pV}{(2\pi )^3} (E_p-\mu ) - 12\int_{regul}\frac{d^3pV}{(2\pi )^3} (E_p+\mu ) + C .
\end{eqnarray}
The 1st and 2nd line of (\ref{fun}) describe the spectrum of quasiparticles constructed from the quarks 
($\alpha_{a}, \beta_{a}$) and antiquarks ($\tilde{\alpha}_{a}, \tilde{\beta}_{a} $).
An example of such a spectrum is pictured in the Fig. 1 as a function of $|\vec{p}|$. The
lower branch describes the "quark" and the higher "antiquark" quasiparticle dispersion relations. 
The minimum of the "quark" spectrum coincides with the Fermi energy ($\mu =0.38$ GeV in the Fig. 1) 
and its value is equall to the size of the energy gap. The low energy exictations around the Fermi surface thus consist
of the "quark" quasiparticles and the contribution from antiquarks can be neglected. This is the main
reason why in the first approximation the relativistic theory behaves very much like the non-relativistic field theory
of condensed matter system. Due to the Pauli principle the pair creations are impotrant only at the energy scales of 
the order of the double chemical potential $\mu $ or in the subtle effects like the decay of the "pion"\cite{nowak}.
The 3rd line of (\ref{fun})
describes the contribution from unpaired "Blue" quarks and antiquarks. The operator $c^{j}_{B, s}$ is defined by the following
"Bogoliubov transformation":
\begin{equation}
c_{B,s}^j=\left\{\begin{array}{lr}
a_{B,s}^{j\,\dagger } & E_p<\mu \\
a_{B,s}^j & E_p > \mu \\
\end{array}\right. .
\end{equation}
The operator part of the hamiltonian (\ref{fun}) is positively defined. This let us
to define the vacuum state $|\Phi\rangle $ as the "trivial" with respect to the new basis operators:
\begin{equation}
\alpha_a|\Phi\rangle =\beta_a|\Phi\rangle =\tilde{\alpha}_a|\Phi\rangle = \tilde{\beta}_a|\Phi\rangle =
c_{B,s}^j|\Phi\rangle =b_{B,s}^j|\Phi\rangle =0
\end{equation}
The c-number part of the hamiltonian (\ref{fun}) gives the vacuum energy. In the 5th line we have the contribution 
from pairing interactions of quarks and the last line describes the contribution from the Fermi sea 
of "Blue" quarks and from the Dirac sea (the last term) of all antiquarks. 
The Dirac sea term is only one divergent in (\ref{fun}). In that place we used additional cut-off regularization
which was discussed before. The constant $C$ is choosen such that the preassure and number of particles at zero
density is zero. The gap equations for $M$ and $|\Delta |$ is obtained by the variation of 
the vacuum expectation value of  (\ref{fun}) with respect to $M$ and $\Delta $.

\begin{figure}
\centerline{\epsfxsize=7 cm \epsfbox{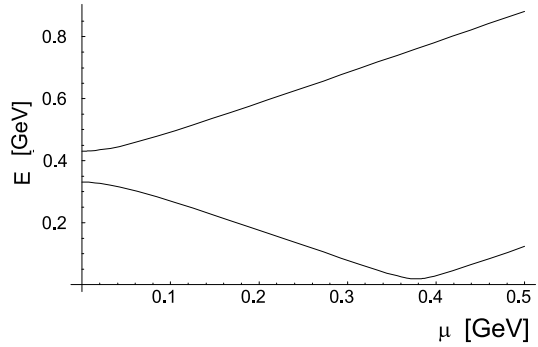}}
\caption{Spectrum of quasiparticles at $\mu = 0.38$ GeV for massive quarks $m=7$ MeV and
$G=G^{\,\prime }=5.01$ GeV$^{-2}$}
\end{figure}

\section{Results}

Let us first consider the case of massless quarks $m=0$.
The values of $G=5.01$ GeV$^{-2}$ and $\Lambda = 0.65$ GeV can be fitted to the values of the chiral condensate
and the pion decay constant \cite{klev}. This choice of parameters establishes that the chiral phase transition is first order,
its position is at $\mu = 0.316$ GeV, and the size of constituent quark mass at the transition point is $M=0.29$ GeV. 
The values of $G^{\,\prime }, \omega $ have little influence on these results. On the other hand,
the position of the superconducting phase transition (CS) and the size of diquark condensate $\Delta $ do depend on these parameters.
The value of $\omega = 0.8$ GeV has been chosen arbitrarily and the behavior of CS phase transition has been checked as a
function of $G^{\,\prime }$. For low values (numerically below $G/2$) the chiral and CS phases are separated in
the density phase diagram. At higher values of $G^{\,\prime }$ there are coexistence regions. Let us consider two cases
$G^{\,\prime }=G/4$ and $G^{\,\prime }=G$.
For $G^{\,\prime }=G/4$ the chiral phase transition is placed at $\mu = 0.316$ GeV and then at $\mu = 0.7$ GeV only a trace of 
continuous CS phase transition appears (in a sense that the value of $\Delta $ reach roughly 1 MeV at best and
quickly vanishes). The vanishing of $\Delta $ at high density is a general feature of the model and is discussed
at the end of this section.
For $G^{\,\prime } = G$ another scenario is realized.
First at around $\mu = 0.305$ GeV there appears a continuous phase transition to the CS phase 
at non-zero $M=0.301$ GeV. The gap $\Delta $ develops up to the value of 3 MeV at $\mu\approx 0.317$ GeV where there is 
the first order chiral phase transition. The gap parameter $\Delta $ changes from about 3 MeV to 18 MeV. The
size of the gap $\Delta $ is not very high in compare to e.g. \cite{cs2} however it size depends on the values
of the parameters $G^{\,\prime }, \omega $. Increasing the density increases the size of $\Delta $.
It reaches a maximum value of $\Delta =43$ MeV at $\mu = 0.55$ GeV.

In the real world the chiral symmetry is explicitly broken by current masses of quarks. For
$u, d$ quarks we take them equal, $m= m_u=m_d=7$ MeV. Inclusion of massive quarks leads to
a little bit higher values of the gap parameter $\Delta $ in the coexistence phase compared to the massless case and changes
the position of phase transitions in the phase diagram. As an example let us consider the
case of $G^{\,\prime } = G$. The phase diagram is given in the Fig. 2. 
It can be seen that at $\mu\approx 0.33$ GeV we have a continuous superconducting phase transition.
The value of the gap $\Delta $ increases with increasing density up to about 8 MeV at $\mu\approx 0.341$ GeV.
Then there appears a first order chiral phase transition and parameters change from $M=0.29$ GeV to
$M=0.15$ GeV and $\Delta = 8$ MeV to $\Delta= 20$ MeV (in the Fig. 3 there is the contour plot
of the potential as a function of $M,\Delta $ at $\mu = 0.341$ GeV). When the density is further increased the value
of $M$ decreases and $\Delta $ increases with maximum around 43 MeV at $\mu\approx 0.56$ GeV.

For both massive and massless quark cases at high enough densities ($\mu $ around 0.9 GeV) the gap
$\Delta $ vanishes. This comes as a result of the regularization by the form-factor. At high densities 
the energy of interaction is too small to overcome the volume energy of the creation
of the superconducting condensate. In this region the superconducting gap
has to be calculated through the use of another approximation.

In conclusion, one can find that within the NJL model, using the relation $G^{\,\prime }=G/4$ between
diquark and quark-antiquark channels coupling constants, there is no superconducting phase transition
in the region of model applicability.  However, when one releases this constraint, and varies the value
of $G^{\,\prime }$, there are rich structures of the  chiral and superconducting
phase transitions. For lower values of $G^{\,\prime }$, these transitions are separated in density phase diagram, 
whereas for
higher values there is a coexistence region. The non-zero current quark masses has only minor influences
for the whole picture.

\begin{figure}
\centerline{\epsfxsize=7 cm \epsfbox{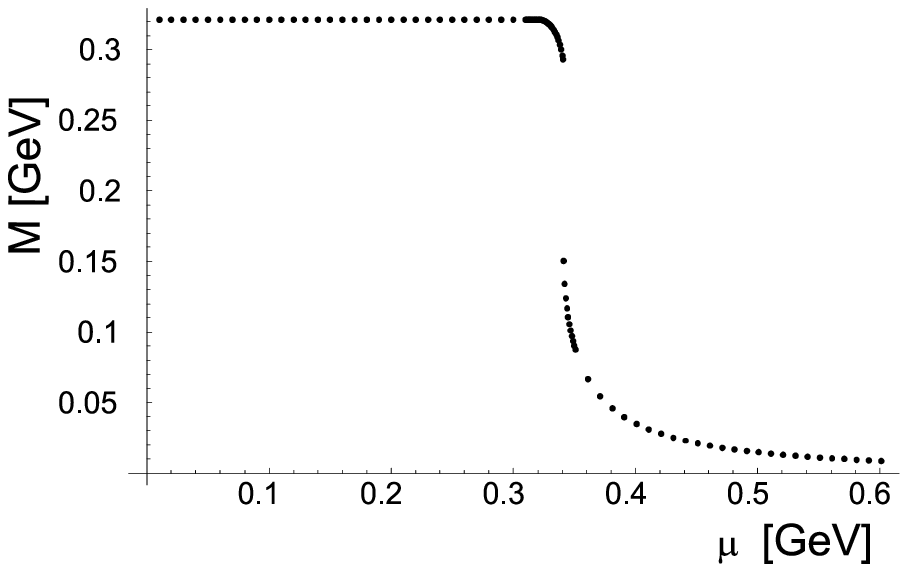} \epsfxsize=7 cm \epsfbox{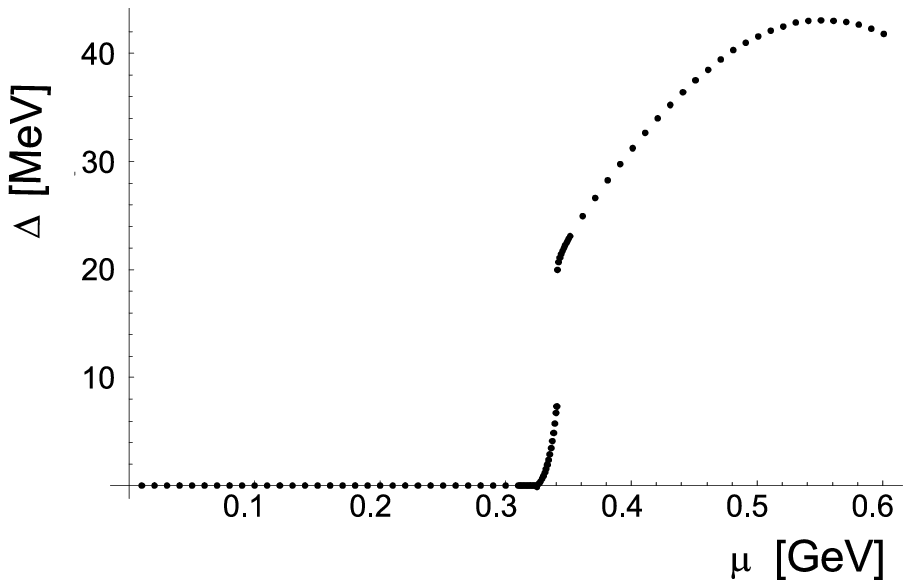}}
\caption{Dependence of $M$ and $\Delta $ on chemical potential $\mu $}
\end{figure}

\begin{figure}
\centerline{\epsfxsize=7 cm \epsfbox{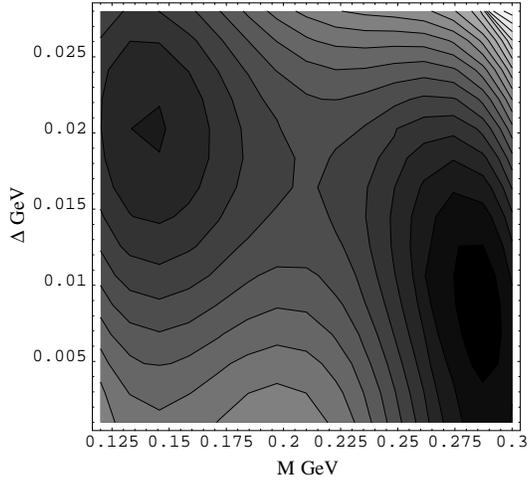}}
\caption{Contour plot of free energy as a function of $M$ and $\Delta $ at the vicinity of the phase transition at $\mu = 0.341$ GeV.}
\end{figure}

\bigskip

\noindent
{\bf Acknowledgement} I would like to thank Jozef Spa\l{}ek, Wojtek Broniowski and
Leszek Hadasz for useful discussions. This work was supported by 
Polish State Committee for Scientific Reseaerch, grant no. 2P 03B 094 19.

\end{document}